# Synergizing Human Expertise and AI Efficiency with Language Model for Microscopy Operation and Automated Experiment Design

Yongtao Liu,[1*] Marti Checa,[1] Rama K. Vasudevan[1]

[1] Center for Nanophase Materials Sciences, Oak Ridge National Laboratory, Oak Ridge, Tennessee, USA

*Corresponding: liuy3@ornl.gov

**Abstract**

With the advent of large language models (LLMs), in both the open source and proprietary domains, attention is turning to how to exploit such artificial intelligence (AI) systems in assisting complex scientific tasks, such as material synthesis, characterization, analysis and discovery. Here, we explore the utility of LLM, particularly ChatGPT4, in combination with application program interfaces (APIs) in tasks of experimental design, programming workflows, and data analysis in scanning probe microscopy, using both in-house developed API and API given by a commercial vendor for instrument control. We find that the LLM can be especially useful in converting ideations of experimental workflows to executable code on microscope APIs. Beyond code generation, we find that the GPT4 is capable of analyzing microscopy images in a generic sense. At the same time, we find that GPT4 suffers from inability to extend beyond basic analyses or more in-depth technical experimental design. We argue that a LLM specifically fine-tuned for individual scientific domains can potentially be a better language interface for converting scientific ideations from human experts to executable workflows, such a synergy between human expertise and LLM efficiency in experimentation can open new door for accelerating scientific research, enabling effective experimental protocols archive and sharing in scientific community.

Recent advancements in large-language models (LLMs) have demonstrated exceptional performance in natural language processing tasks.[1-4] The state-of-the-art LLMs also show potential in analyzing and generating text related to scientific disciplines at a high level of sophistication.[5-8] This garnered widespread interest across various scientific fields, such as chemistry, materials science and biology. For example, recent studies showed that LLMs facilitate molecule and property predictions[9-12] knowledge extraction and database enhancement[13-17], research reproduction[18], and education applications[19].

Beyond language processing, LLMs are also increasingly adept at interpreting and writing code. Nowadays, there is also a growing trend of incorporating application program interfaces (APIs) in scientific instruments, allowing users to control instruments programmatically for automated experimentation (AE). These open new avenues to leverage LLMs in experiments by chaining the LLMs with APIs to assist researchers in experiment design, tool operation, and data analysis. The potential of LLMs to transform human language, e.g., demonstration of experiments or scientific questions, into workflow scripts that can be executed in instruments through APIs is particularly promising for the application of LLMs in assisting operations of scientific instruments,[20] as well as data analysis.

This study explores the role of LLMs as a language interface in scientific instrumentation, with a focus on microscopy which has been the cornerstone of nanoscale characterizations and manipulation for over three decades.[21,22] Traditionally, mastering microscopy techniques for scientific discovery often requires years of training. Recently, the rapid progress in the artificial intelligence (AI) field propelled the development of AE in microscopes for physics discoveries.[23-28] AE-microscopes utilize a range of machine learning (ML) methods, including supervised ML,[24,29] active learning,[27,30-34] and reinforcement learning[26,28,35,36] for on-the-fly data analysis and decision-making, automating microscope operation, and facilitating physics discovery. For example, convolutional neural networks (CNN) can be used to transform streaming microscope images into segmented images highlighting objects of interest, (e.g., convert a topography to a grain boundary map, piezoresponse force microscopy (PFM) amplitude to a map of domain wall locations, a scanning transmission electron microscopy (STEM) to atomic coordinates, etc.).[24,29] This allows systematic investigation of the objects of interest through spectroscopic measurements at the segmented objects or manipulation at nano and atomic scale. Similarly, object detection algorithms utilizing deep learning, such as YOLOv3, can provide the locations of molecules on captured images, enabling high-resolution imaging without user intervention.[37]

Given the increasing interest and application of AE,[23,38-40] designing and programing AE workflows become a required skill in the field. A workflow includes a range of elements such as operating actions, data analyses, and decision-making for an experimentation.[41] Traditionally, workflows are often rooted in the mind of experienced operators and not documented prior to experiments. During experiments, operators have the chance to fine tune the sequence of specific elements to realize the experimentation ideas. However, conducting a workflow in an automated manner requires prior programming, and must include all experimental elements in the correct sequence. As such, researchers in the AE field need to expend considerable effort in converting experiment ideas into workflow scripts that can be executed on the instrument. We suggest that this bottleneck can be potentially addressed via LLMs, which can be used to analyze demonstrations of experimentation ideas and convert them into efficient and reproducible workflows that can be executed through instrumentation APIs.

In this work, we investigate the potential of combining LLMs with APIs to serve as a language interface assisting researchers in preparing workflows and operating scanning probe microscopy (SPM). We explore the LLMs in various scenarios including text-to-program conversion, programming workflow, reproducing experiments, and data analysis. We illustrate the application of LLMs in both custom APIs and commercialized APIs. The results suggest that the integration of LLMs and APIs enables more efficient and reproducible experiments, which can potentially significantly increase the pace of scientific research. The integration of LLMs and APIs for archiving experimentation can also serve for FAIR infrastructure development.[42] The approach can further extend to integrate speech-to-text models for voice control of scientific instruments.

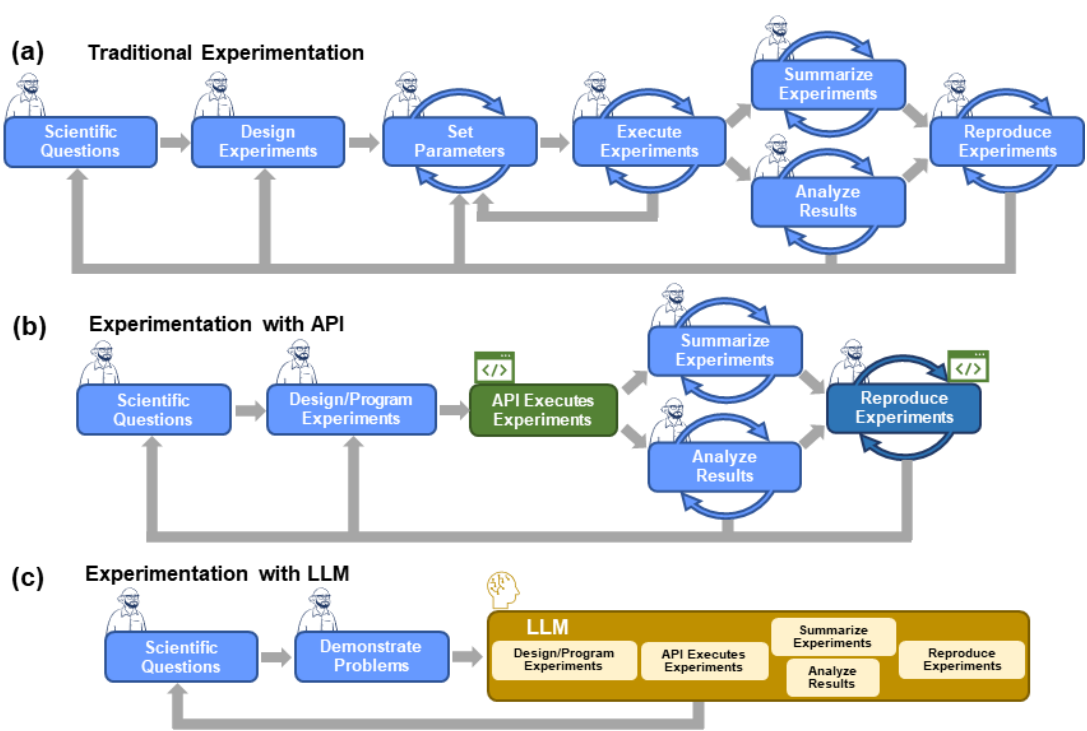

**Figure 1.** Processes of (a) traditional experimentation, (b) API-assisted experimentation, and (c) LLM-assisted experimentation. API-assisted experimentation reduces the need for repetitive, manual operation by human researchers. LLM-assisted experimentation further reduces the required efforts of programming workflows.

Traditional experimentation typically includes repetitive tasks such as setting experimental parameters, executing experiments, summarizing experiments and findings, and analyzing results, as schematically shown in Figure 1a. These tasks often require certain level of expertise and manual operation by human researchers. Also, the reproduction of experiments, whether by the original researchers or others, necessitates repetition of the exact processes. The integration of APIs can change this paradigm by enabling users to directly execute the workflow script of specific experimentations through APIs (Figure 1b), reducing the need for manual and repetitive operations. Furthermore, these workflow scripts can be archived and reused for the reproduction of experiments in the future. However, the utilization of APIs still demands specific skills, such as programming expertise and the ability to design effective workflows, which are not inherently intuitive for all researchers. Given the ability of analyzing text and writing code from human instructions, as well as interpreting existing code, LLMs have the potential to function as language interface to assist programming in API environment. This language interface would allow for converting human instructions to executable workflow scripts, assisting in data analysis, and automating the reproduction of experiments, potentially democratizing access to AE and streamlining research processes (Figure 1c).

**Training LLM with Instrumentation Knowledge**

We employ ChatGPT4 and AEcroscoPy to investigate the application of LLMs in API environment. ChatGPT4 is an advanced model in the LLM landscape.[1,4] AEcroscoPy is a custom-developed API, a cross-platform software-hardware system designed for microscope automation.[43] For readers interested in AEcroscoPy, we refer to the recent publication ref[43] where details of AEcroscoPy are extensively discussed. We feed a complete guideline[44] of AEcroscoPy into ChatGPT4 as a learning resource. This guideline consists of a thorough overview of conducting SPM experiments using AEcroscoPy, including aspects such as installation procedures, software initialization, and a range of functionalities. These functionalities are tailored for operating a SPM via Python scripts, including a spectrum of tasks: setting input/output parameters, moving the SPM tip, conducting raster scan imaging, conducting point or grid spectroscopy measurements, customizing scan trajectories, and more. The feedback from ChatGPT4 (Figure 2) indicates a good assimilation of the key components in the AEcroscoPy guideline. Hereinafter, we refer to the ChatGPT4 learned AEcroscoPy guideline as AE-GPT and use it for further investigation.

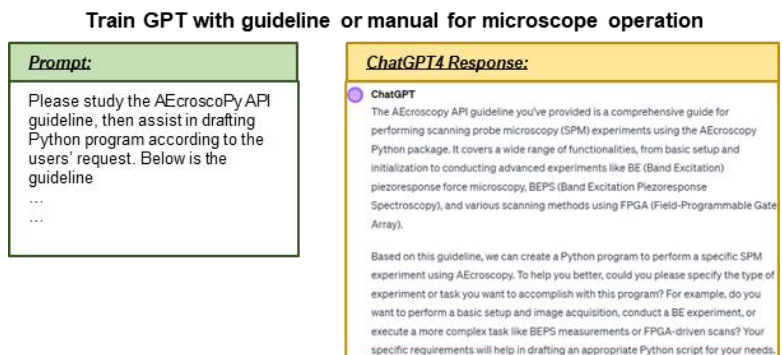

**Figure 2.** Train ChatGPT4. The AEcroscoPy guideline is feed into ChatGPT4 as the learning resource, ChatGPT4's response indicates that it learned the key components in AEcroscoPy. The trained ChatGPT4 is referred as AE-GPT and will be used for further study.

**Converting Instructions to Executable Scripts**

In the initial examination, we focused on assessing AE-GPT's proficiency in converting individual human commands for SPM operation into Python script. We first query AE-GPT about the method of setting parameters and moving the SPM tip. AE-GPT correctly suggested the appropriate function, *tip_control()*, to use, as shown in Figure S1. Further examination is to query about setting band excitation (BE) amplitude and center frequency to 380 kHz. AE-GPT also provided the correct program using *define_be_parms()* function. Notably, we only specified two of seven parameters in the *define_be_parms()* function, in this case, AE-GPT opts to use the default value for other five parameters.

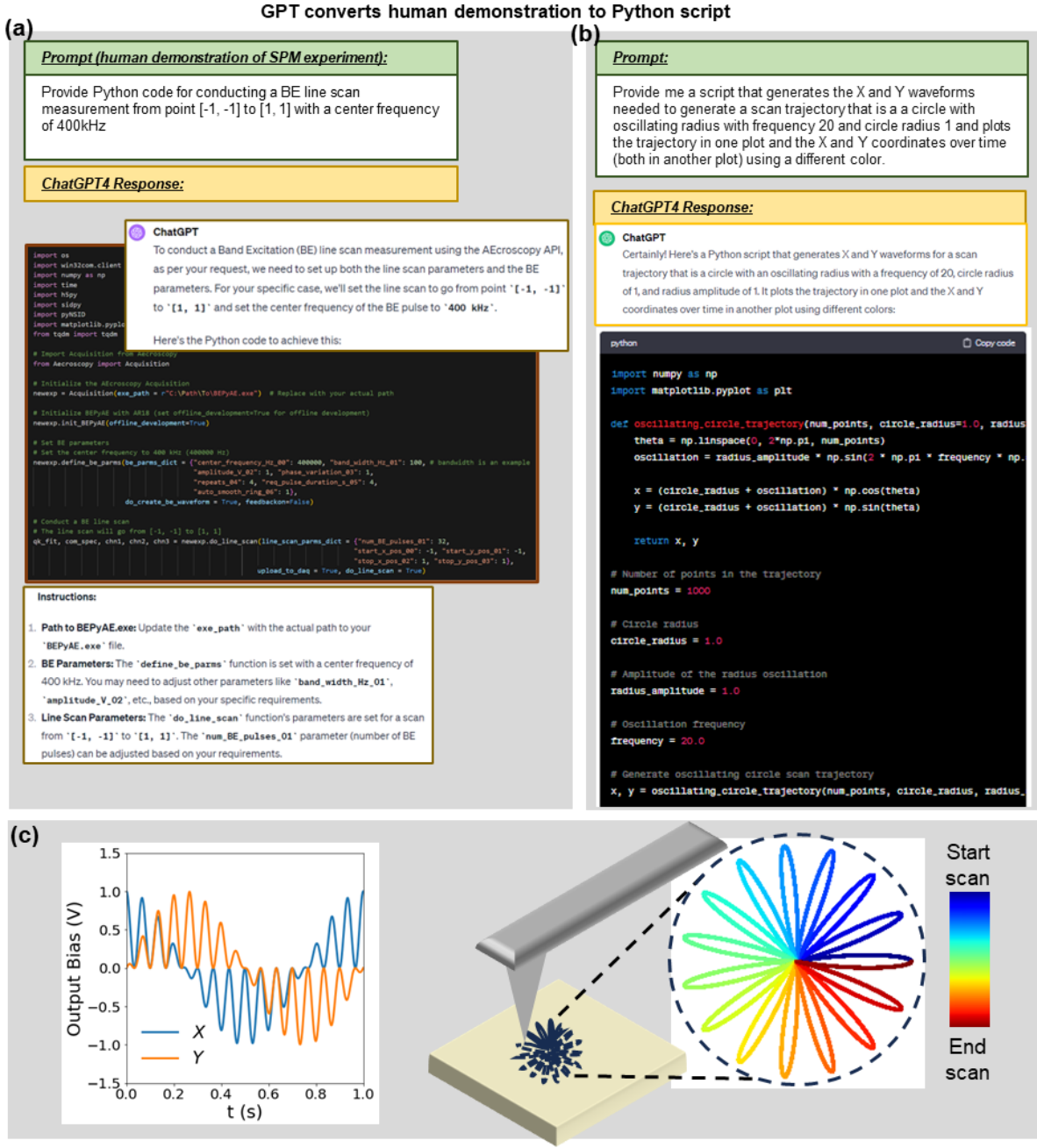

**Figure 3.** Examination of converting individual command to Python code using AE-GPT. (a) Here the command is to perform a BE line scan with defined start and end points. AE-GPT provided a comprehensive code snippet including importing packages, initializing software, and performing

required tasks. (b) AE-GPT design of XY tip trajectory for an arbitrary scan in flower-shaped mode. (c) X, Y output waveforms and final scan trajectory.

Building on this success, we proceed to examine the ability of AE-GPT in programming an entire measurement workflow that integrates multiple functions in a specific sequence. Programming an entire measurement workflow is a task that demands considerable workflow design skills. To test this, we asked AE-GPT to program a BE line scan measurement, specifying scan start/end points and BE center frequency. In practice, using AEcroscoPy for this measurement includes two steps: first is to set the BE parameters, second is to set line scan coordinates and trigger line scan measurements. AE-GPT correctly utilized *define_be_parms()* and *do_line_scan()* functions in sequence to fulfill these objectives, as shown in Figure 3a. In the process, AE-GPT also successfully imported the required class *('Acquisition')* from AEcroscoPy and initialized the software.

We further presented AE-GPT with a more challenging task: conduct a spiral scan with a 5V voltage applied to the AFM tip. Our AEcroscoPy incorporates a Field Programmable Gate Arrays (FPGA) to enable customized scan trajectories, such as a spiral scan. While the AEcroscoPy guideline includes instructions for performing a spiral scan, integrating a DC voltage application to the tip is nontrivial. For AE-GPT, the difficulty of applying a DC voltage arises from the necessity of understanding not only the functionalities of the AEcroscoPy package but also the specific hardware configurations. However, there is not much hardware knowledge in the learning source provided to AE-GPT, making this task particularly challenging. Nonetheless, AE-GPT's response remains insightful, as shown in Figure S2. It identified the relevant functions necessitated to the experiment. The flaw in AE-GPT's response is that it erroneously includes the tip voltage parameters directly into the spiral function, which does not align with the actual operation of AEcroscoPy. We would attribute this misstep the absence of necessary information in the learning source provided to AE-GPT. Nonetheless, AE-GPT's overall analysis of the inquiry and the approach to devising a solution are good, highlighting its potential in understanding and addressing experimental setups. Notably, AE-GPT's response not only include program code but also comprehensive instruction of the program, allowing users to easily understand the program and fine tune parameters. These examinations demonstrate the proficiency of AE-GPT in learning the guideline and providing relevant program according to user inquiries.

An additional feature that AEcroscoPy package provides is to design custom tip scan trajectories, which allows to overcome some limitation of classical raster-scan based tip paths such as scan speed, tip and/or sample wear, or specific lithographic scan paths, among others.[45,46] However, the intricate aspect of designing a custom trajectory is to determine the X, Y waveforms that will drive the probe movement through the piezo positioners. AE-GPT can also help in this challenging task as it is depicted in Figure 3b-c. In Figure 3b a, we show an example of the prompt a human user would need to provide to ChatGPT4 and its subsequent response to generate the signals that the SPM controller needs to perform an arbitrary scan shaped like a circle with oscillating radius (Figure 3c). Such type scans can be helpful in the writing of certain lithographic patterns in ferroelectric materials and are not accessible with most commercial SPM systems. The last step, as explained in previous sections, is to route the generated X,Y waveforms to the X,Y piezo positioners of the SPM controller (Figure 3c). Following this methodology, any arbitrary design of a scan path can be easily generated and added in the AE code.

**Programming Multifaceted Workflow**

In further exploration of AE-GPT's capabilities, we tested its proficiency in programming an experiment that integrates several measurements. The presented experiment workflow comprises the following steps:

    (a) perform a BE raster scan.
    (b) identify and extract domain wall locations from PFM results.
    (c) conduct BEPS (Band Excitation Piezoresponse Spectroscopy) measurements at domain walls.
    (d) apply a DC pulse at the domain walls and re-image to observe changes.

Figure 4 shows the process of this exploration, detailed prompts and AE-GPT's responses are shown in Figure S3. After analyzing the workflow, AE-GPT pointed out that such workflow needs custom functions that extend beyond standard API functions: "*This kind of workflow typically requires custom scripting beyond standard API functions, especially for steps like identifying domain wall locations from PFM results, which usually involve advanced image processing techniques*". This is absolute for tasks like identifying domain wall locations from PFM images, which often necessitates specific image processing approaches.[24,29,47] AE-GPT proceeded to design a conceptual workflow script that incorporated the *raster_scan()* function for BE raster scan and *do_beps_specific()* function for BEPS measurement at targeted locations (i.e., domain wall locations). AEcroscoPy package includes two functions for BEPS measurement, one is to do grid locations BEPS and the other is to do BEPS at specific locations, AE-GPT successfully identified that the *do_beps_specific()* is the appropriate function to use here.

The returned workflow also includes a placeholder for the domain wall identification process and AE-GPT suggested a peak detection method for this task. The suggested method sounds reasonable if the domain walls are very sparsely distributed. As a follow up to fill in the placeholder, we prompted AE-GPT to use an edge detector for wall detection, which is often an effective method to detect this kind of features in image data. Then, AE-GPT adapted the script to utilize a Canny filter for domain wall detection (Figure S4). Interestingly, Canny filter was also employed in our previous work to identify domain walls from PFM images.[48]

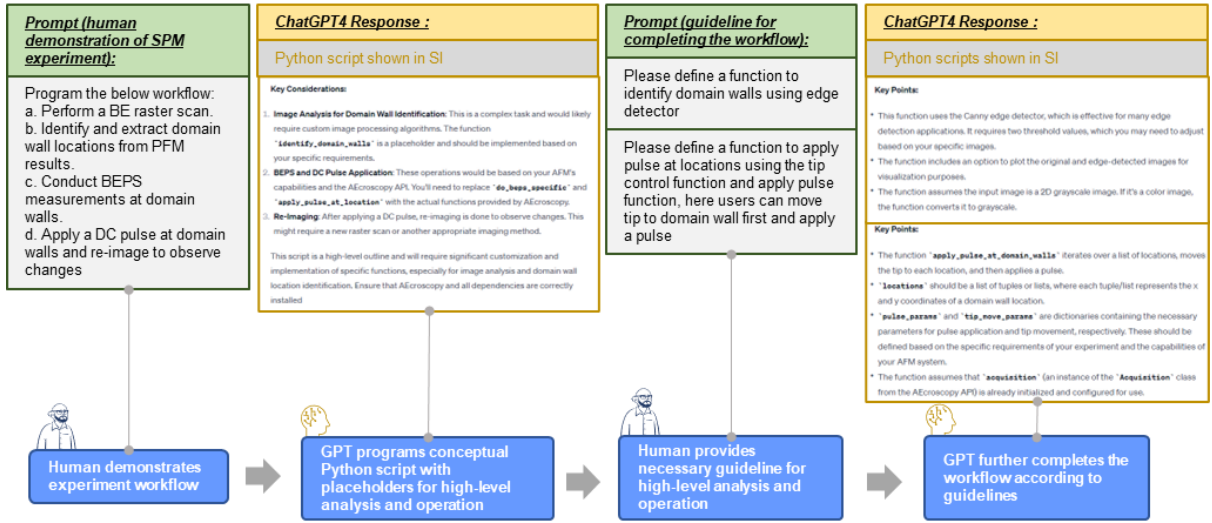

**Figure 4.** Human-LLM collaboration for the design of workflows integrating multiple distinct measurements and tasks.

For step (d) of the above experimental workflow, AE-GPT suggested using a function to apply a pulse at a specific location, which is partially correct in the context of AEcroscoPy. Executing this action in AEcroscoPy actually includes two sub-steps: first, position the AFM tip at the domain wall location; second, apply the pulse. Successfully performing this step necessitates domain knowledge related to microscope operations, which can vary depending on the microscope in use. For instance, some microscope control software and/or API allows independent control of tip movement and separated functions for follow-up actions (e.g., spectroscopic measurement or pulse application)—similar to our AEcroscoPy design. In contrast, other systems integrate tip movement as a sub-function within spectroscopic measurement processes—similar to the approach proposed by AE-GPT. Therefore, the approach proposed by AE-GPT is reasonable even if it does not align with AEcroscoPy's design. To design a function for applying pulse at domain walls using AEcroscoPy, we guided AE-GPT to utilize AEcroscoPy's functions for tip control and pulse application to conduct step (4). With this guide, AE-GPT successfully write the correct script (as shown in Figure S5).

AE-GPT also shows ability of experimental designs for addressing specific scientific questions. For instance, when prompted with a query about PFM experiments for investigation of ferroelectric domain wall dynamics, AE-GPT returned with three experiment designs (Figure S6) focusing on domain wall manipulation, frequency-dependent wall response, and time-resolved wall dynamics. Each approach represents an insightful aspects of ferroelectric domain wall behaviors.

This examination demonstrates that AE-GPT can architect intricate workflow scripts integrating multiple measurement and analysis steps. More importantly, it reveals AE-GPT's proficiency in formulating solutions with human guidance. This fusion of human expertise and the capabilities of LLM implies promising new pathways for collaborative experimental design, where human insight and AI efficiency can synergistically enhance the scientific process.

**Summarizing and Reproducing Experiments**

In research activities, summarizing experimental details in natural language after experiments is crucial for clarity and effective communication between researchers in scientific community. Traditionally, this has been done by researchers based on entries in laboratory notebooks. These approaches rely almost exclusively on the researcher to document all necessary details and are thus liable to miss significant details for reproducing experiments. Thus, in addition to its ability to convert natural language into executable code, it is also curious to explore the ability of AE-GPT in summarizing experiment protocols from codes and/or records into natural language. Our AEcroscoPy package includes a logging feature, which records every executed step and corresponding detailed parameters of an experiment. The experimental details can be saved as a logger file at the end of the experiment. We utilize the logging feature to test the ability of AE-GPT in converting experimental details to a natural language summary. We fed a logger file to AE-GPT, this logger file represents an experiment comprising tip setting, IO cluster setting, and BE parameters setting. AE-GPT successfully translated the logger file into natural language descriptions of these settings. Conversely, when requested to reproduce the experiment detailed in the logger file, AE-GPT is also able to convert the experimental details in the logger file to a comprehensive Python script for the experiment. This examination suggests that AE-GPT not only can summarize experiment but also reproduce previous experiment based on recorded experimental details.

In addition to reproducing experiments conducted by ourselves previously, researchers are also often interested in reproducing experiment reported in literatures. Reproducing experiments described in scientific literature often requires collecting various experimental parameters and conditions, and setting up instruments accordingly. In this context, AE-GPT's ability in analyzing natural language and extracting information can assist. We presented a scientific paper[49] that includes several PFM experiments to AE-GPT, one of the PFM experiments detailed in the paper is BE-PFM. We ask AE-GPT to extract the specific BE-PFM parameters from the paper, which is successfully accomplished by AE-GPT. Building on this, we further requested AE-GPT to write a Python script for reproducing the BE-PFM experiment using AEcroscoPy. AE-GPT offered a script to set the parameters and conduct the experiment. This showcases AE-GPT's ability in translating experimental details from previously published work and practically reproducing them.

The ability in summarizing experimental details and reproducing previous experiments can greatly benefit the scientific community in terms of communication and reproducibility of experiments.

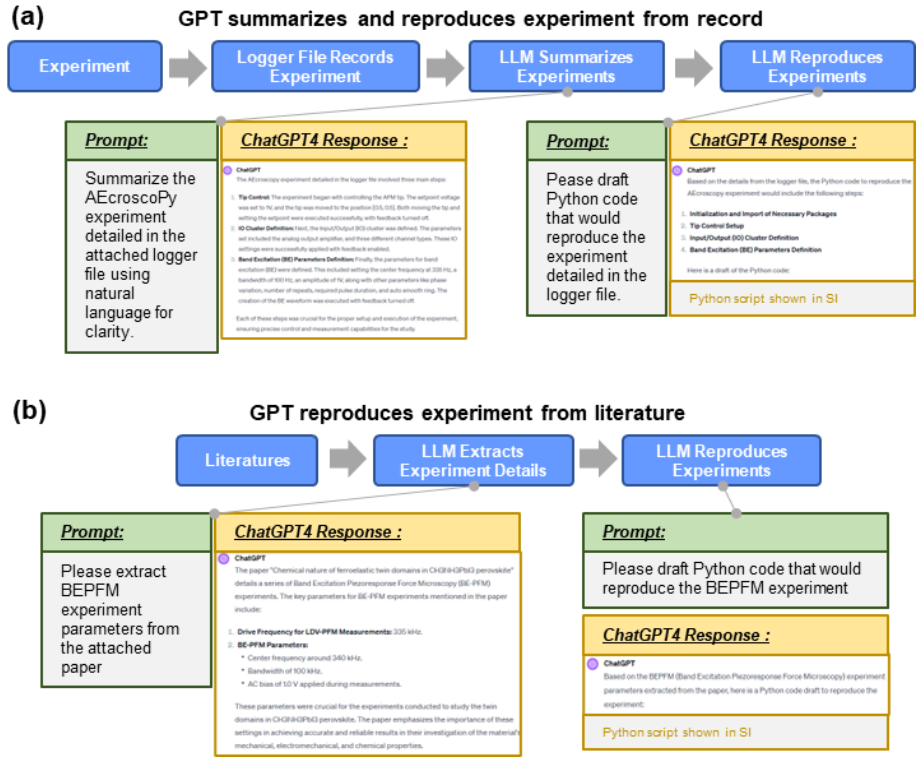

**Figure 5.** Summarize experimental details in natural language and reproduce experiments. (a) The process of LLM summarizes experiment detailed in logger file and reproduces the recorded experiment, the Python script generated by AE-GPT to reproduce the experiment in logger file is shown in Figure S7. (b) The process of LLM extracts experiment detailed in published paper and reproduce it, the Python script generated by AE-GPT to reproduce the experiment in the literature is shown in Figure S8.

**Data Analysis**

In addition to assistance in conducting experiments, we also evaluate how effectively AE-GPT can analyze experimental results. To this evaluation, we feed AE-GPT a BEPFM result that is filed in a Hierarchical Data Format (HDF) comprising both experimental parameters (e.g., drive frequency, excitation waveform, etc.) and experimental results (e.g., amplitude and phase signals, topography, etc.). We sent several inquiries to AE-GPT according to the target information we would like to see. As detailed in Figure 6, AE-GPT can identify experimental parameters, plot raw data and image data upon request. Noteworthily, AE-GPT not only shows the results when asked to plot specific results, but also provides the corresponding Python script with analysis instructions, as shown in Figure S9.

Beyond merely extracting parameters and plotting results, we also asked AE-GPT to perform data analysis, such as "find the spectroscopy with the strongest intensity", and "calculating roughness." As shown in Figure 6, AE-GPT also successfully accomplished these tasks, showcasing its analytical ability. Moreover, we also evaluated the ability of AE-GPT in performing scientific interpretation of the data.

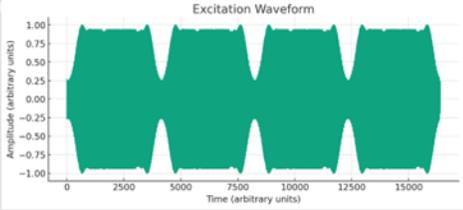

**Figure 6.** Analyzing BEPFM data using GPT.

**Necessity of Science-LLM for Scientific Interpretation**

However, we found that AE-GPT's ability in scientific interpretation is very limited, as shown in Supplemental Note-I. The interpretation is mostly just the definition of certain scientific concept, such interpretation offers limited insights and even can be wrong in some scenarios (e.g., the interpretation of phase variation in a few datasets in in Supplemental Note-I). In addition, AE-GPT is also lack of designing advanced experiments based on literatures, as indicated in Supplemental Note-II, the designed experiments do not harness knowledge from the provided literatures. This is probably because that disentangling scientific information in such a dataset or literatures requires some deeper analysis with more specific analysis approaches and resources, or the absence of critical knowledge in AE-GPT. In this regard, it is expected that LLMs more specifically trained on the scientific literature would be significantly better at these tasks.[50]

**The Potential of Utilizing LLM in a Wide Range of APIs**

We also extended our investigation to two different commercially available APIs, namely, the Nanosurf AFM and the Zurich Instruments lockin amplifier, to test its performance outside of our custom API. Therefore, we trained the ChatGPT4 with Nanosurf Programmer Script Manual and the LabOne Zurich Instruments User and Programming manual, respectively. Subsequently, we investigated its proficiency in translating natural language instructions into codes and writing workflow scripts. These investigations are shown in Supplemental Note-III and Supplemental Note-IV, respectively, where the potential of using LLMs outputs previously trained with commercially available instruction manuals is shown. Despite this broader scope of the guideline, ChatGPT4 also learned from the manual and severed as a script assistant to provide useful scripts upon requests. This indicates that LLMs' potential as a language interface for interacting with different types of scientific instrumentation software to assist in programming experiment and data analysis scripts (Figure 7).

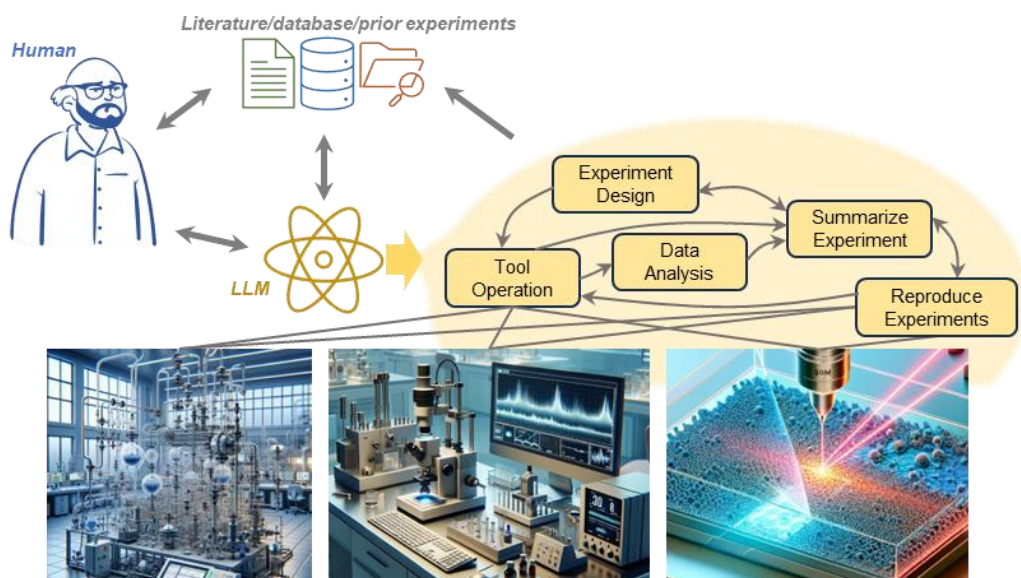

**Figure 7.** The potential of LLM serving as a language interface synergizing human expertise and AI efficiency to facilitate experiment design, execution, and data analysis. The images in this figures are generated by DALLE.

**Conclusions**

In summary, we explored the potential of integrating LLMs into the realm of scientific experimentation operation and data analysis. It demonstrates that combining LLMs with API in scientific instrumentation for experimentation and data analysis promises significant advancement in streamlining the process of experiment design and execution, as well as data analysis. Notable aspects of LLMs' utility are its ability to assist in programming experimentation workflow, archiving experiment process, and the reproduction of experiments from both experimental notes and scientific literature. With these capabilities, the synergy between human expertise and LLM efficiency (Figure 7) in experimentation will open new door for accelerating scientific research, as well as serve for FAIR infrastructure enabling effective sharing experimental protocols and results in scientific community.

We also note that the experimental design capabilities of AE-GPT are generic, and the data analysis especially the interpretation of scientific data is basic. To overcome these drawbacks, domain specific LLMs, as well as fine-tuning by domain experts, will be required. Such domain-specific LLMs, trained with scientific literature, scientific instrument manuals, codebases, etc., can potentially be more advanced and useful for researchers.

**Acknowledgements**
This research was supported by the Center for Nanophase Materials Sciences (CNMS), which is a US Department of Energy, Office of Science User Facility at Oak Ridge National Laboratory. Y.L. wishes to express sincere gratitude to Sergei V. Kalinin for helpful discussion.

**Conflict of Interest**
The authors declare no conflict of interest.

**Authors Contribution**
Y.L. conceived the idea. Y.L. performed the investigation. M.C. performed scan trajectory analysis and the Zurich Instruments Lockin section. All authors contributed to discussions and manuscript edit.